\documentclass[showpacs,oneside,twocolumn,prl,amsmath,amssymb,superscriptaddress]{revtex4-1}
\usepackage{cases}
\usepackage{amsmath}
\usepackage{amssymb}
\usepackage{amsfonts}
\usepackage{amssymb}
\usepackage{dcolumn}
\usepackage{bm}
\usepackage{bbm}
\usepackage{graphicx}
\usepackage{xcolor}
\usepackage{array}
\usepackage{subfigure}
\usepackage[none]{hyphenat}

\begin{document}

\title{Mechanical dissipation below 1$~\mu$Hz with a cryogenic diamagnetic-levitated micro-oscillator}
\author{Yingchun Leng}
\affiliation{National Laboratory of Solid State Microstructures and Department of Physics,  Nanjing University, Nanjing 210093, China}

\author{Rui Li}
\affiliation{Hefei National Laboratory for Physical Sciences at the Microscale and Department of Modern Physics, University of Science and Technology of China, Hefei 230026,
China}
\affiliation{CAS Key Laboratory of Microscale Magnetic Resonance, University of Science and Technology of China, Hefei 230026, China}
\affiliation{Synergetic Innovation Center of Quantum Information and Quantum Physics, University of Science and Technology of China, Hefei 230026, China}

\author{Xi Kong}
\affiliation{National Laboratory of Solid State Microstructures and Department of Physics, Nanjing University, Nanjing 210093, China}

\author{Han Xie}
\affiliation{National Laboratory of Solid State Microstructures and Department of Physics, Nanjing University, Nanjing 210093, China}

\author{Di Zheng}
\affiliation{National Laboratory of Solid State Microstructures and Department of Physics, Nanjing University, Nanjing 210093, China}

\author{Peiran Yin}
\affiliation{National Laboratory of Solid State Microstructures and Department of Physics, Nanjing University, Nanjing 210093, China}

\author{Fang Xiong}
\affiliation{National Laboratory of Solid State Microstructures and Department of Physics, Nanjing University, Nanjing 210093, China}

\author{Tong Wu}
\affiliation{National Laboratory of Solid State Microstructures and Department of Physics, Nanjing University, Nanjing 210093, China}

\author{Chang-Kui Duan}
\affiliation{Hefei National Laboratory for Physical Sciences at the Microscale and Department of Modern Physics, University of Science and Technology of China, Hefei 230026,
China}
\affiliation{CAS Key Laboratory of Microscale Magnetic Resonance, University of Science and Technology of China, Hefei 230026, China}
\affiliation{Synergetic Innovation Center of Quantum Information and Quantum Physics, University of Science and Technology of China, Hefei 230026, China}

\author{Youwei Du}
\affiliation{National Laboratory of Solid State Microstructures and Department of Physics, Nanjing University, Nanjing 210093, China}
\affiliation{Collaborative Innovation Center of Advanced Microstructures and Jiangsu Key Laboratory for Nano Technology, Nanjing 210093, China}

\author{Zhang-qi Yin}
\affiliation{Center for Quantum Technology Research and Key Laboratory of Advanced Optoelectronic Quantum Architecture and Measurements (MOE),
School of Physics, Beijing Institute of Technology, Beijing 100081, China}

\author{Pu Huang}
\email{hp@nju.edu.cn}
\affiliation{National Laboratory of Solid State Microstructures and Department of Physics, Nanjing University, Nanjing 210093, China}

\author{Jiangfeng Du}
\email{djf@ustc.edu.cn}
\affiliation{Hefei National Laboratory for Physical Sciences at the Microscale and Department of Modern Physics, University of Science and Technology of China, Hefei 230026,
China}
\affiliation{CAS Key Laboratory of Microscale Magnetic Resonance, University of Science and Technology of China, Hefei 230026, China}
\affiliation{Synergetic Innovation Center of Quantum Information and Quantum Physics, University of Science and Technology of China, Hefei 230026, China}

\begin{abstract}
 Ultralow dissipation plays an important role in sensing applications and exploring macroscopic quantum phenomena using micro-and nano-mechanical systems. We
 report a diamagnetic-levitated micro-mechanical oscillator operating at a low temperature of 3K with measured dissipation as low as $0.59~\mu$Hz and a quality factor as high
 as
 $2\times10^7$. To the best of our knowledge the achieved dissipation is the lowest in micro- and nano-mechanical systems to date, orders of magnitude improvement over
 the reported state-of-the-art systems based on different principles. The cryogenic diamagnetic-levitated oscillator described here is applicable to a wide range of mass,
 making it a good candidate for measuring both force and acceleration with ultra-high sensitivity. By virtue of the naturally existing strong magnetic gradient, this system
 has great potential in quantum spin mechanics study.

\end{abstract}

%\pacs{03.65.Ud, 03.65.Ta,76.70.Hb, 76.30.Mi}

\maketitle
\textit{Introduction.} \textbf{---}Micro- and nano-mechanical systems have been developed as force sensors for detecting a wide range of weak signals, such as charge
\cite{Cleland}, spin \cite{Rugar}, mass \cite{Jensen}, etc. It has also been considered recently as potential tools for exploring force from new physics, such as a variety of
dark matter models \cite{dark mattter-chameleon,dark mattter-gravity detection,dark mattter-Compact Mechanical}, wave-function collapse models  \cite{wave-function
collapse1,wave-function collapse2,wave-function collapse3,wave-function collapse4,wave-function collapse5,wave-function collapse6}, corrections to Newtonian gravity
\cite{Newtonian gravity1,Newtonian gravity2} and the high frequency gravitational wave \cite{High-Frequency Gravitational Waves}. In experimental realization, the signal
understudy should be larger than the noise level of the force sensor that  is characterized by noise power density:
\begin{equation}
\label{force} S_{\rm FF}(\omega_0)=4m \gamma k_{B}T,
\end{equation}
where $\gamma$ is the mechanical dissipation character. One direct way to improve the performance is to make the oscillator mass  $m$ extremely small \cite{trap ions} so as to
achievable the goal even at room temperature. However, for a wide range of applications, such as the force from gravity or acceleration, sufficient large mass is important, so
low temperature environments and the lower mechanical dissipation $\gamma$ are more useful in improving force sensitivity.

\begin{figure*}\centering
\includegraphics[width=1.92\columnwidth]{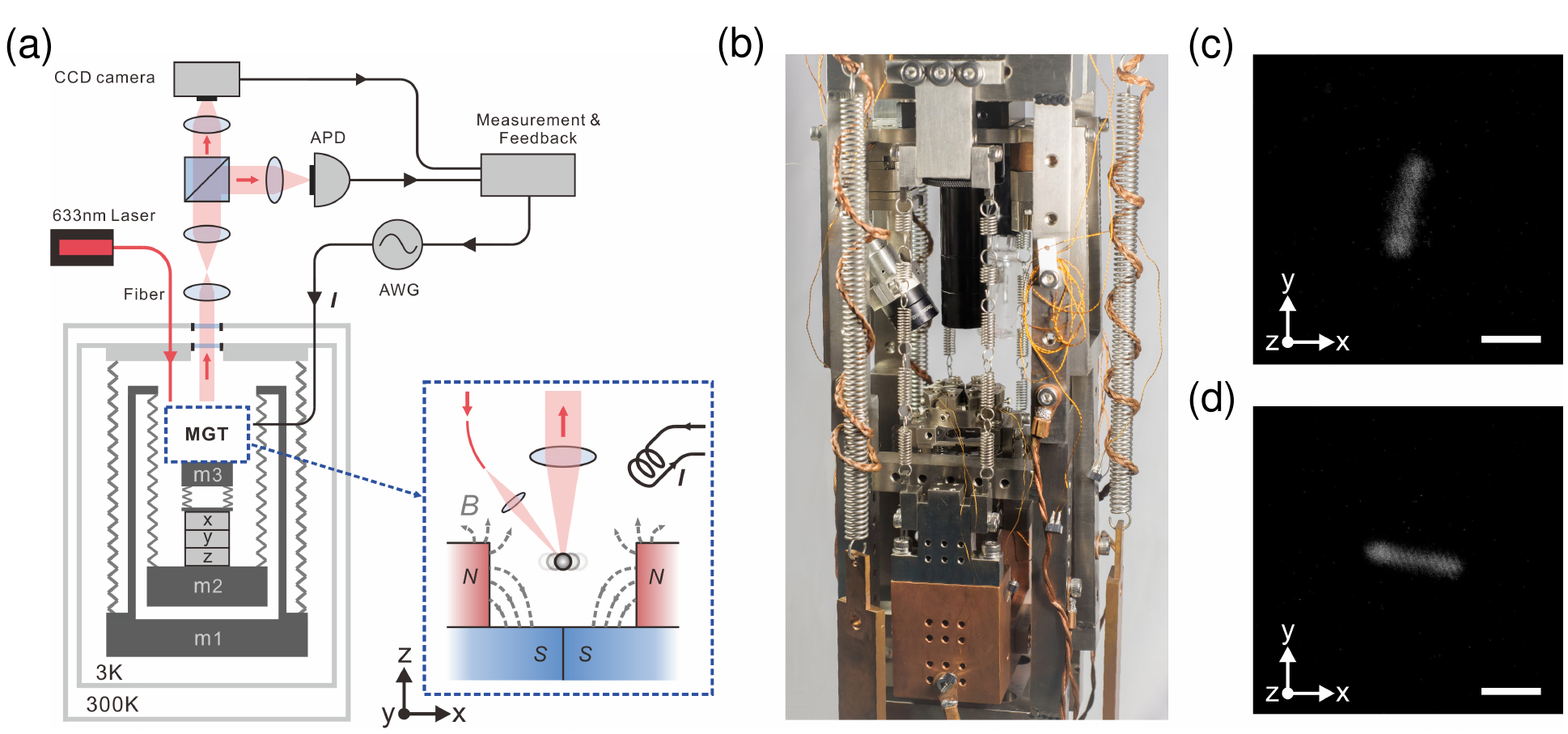}
\caption{(Color online). Experiment setup.
(a), The magneto-gravitational trap (MGT) is placed on the three-stage spring-mass suspension system which is used to isolate external vibration, the whole structure is mounted
inside the chamber of a cryostat, a 633~nm laser light reaches the sample chamber though fiber and is focused loosely on the micro-sphere, and piezo positioners
are used to control the position of trap so that scattered light is collected via lens. A CMOS camera is used to record the position of the micro-sphere, and an avalanche
photodiode detector is used to measured the real-time intensity of the scattered light that is used to feedback cool or excite the motion of the micro-sphere via magnetic force
via a magnetic field generated by a small coil near the MGT.
(b), Vibration isolation system, the picture of spring-mass suspension structure and the MGT placement inside it, especially , soft copper braids are used to realize thermal
connection between.
(c) and (d), snapshots of the excited motion of the micro-sphere corresponding to two oscillation modes of different resonant frequencies (mode 1: 11.7~Hz, mode 2: 8.4~Hz) in
$x$-$y$ plane, where the exposure time is 200~ms, much longer than the period of oscillation (scale bar is 30~$\mu$m).
}
 \label{setup}
\end{figure*}

Besides sensing applications, dissipation is also paramount important for study quantum phenomena at macroscopic scale using micro- and nano-mechanical systems
\cite{Macroscopic quantum mechanics1,Macroscopic quantum mechanics2}. In order to control a macroscopic quantum state, such as generating spatial superposition, an important
requitement is to overcome thermal decoherence. The finite-temperature oscillator's thermal decoherence rate $\gamma_{\rm th} = \overline{n} \gamma$
with $\overline{n}=k_{B}T/\hbar \omega_0 $ the thermal average phonon number \cite{Macroscopic quantum mechanics3}. Especially, in order to maintain quantum coherence within
one oscillation period, we obtain the relation:
\begin{equation}
\label{quantum} \frac{\omega_0^2}{\gamma} > k_{B}T/\hbar.
\end{equation}
The left hand side is equal to the product of oscillator resonance frequency $\omega_0$ and quality factor $Q = \omega_0 / \gamma^{}$, known as $\omega_0 Q$ product
\cite{Macroscopic quantum mechanics3},
so given $\omega_0 $ and environmental temperature $T$, a lower mechanical dissipation $\gamma$ leads to better quantum coherence.

For a solid state mechanical resonator operating  at high vacuum, the dissipation mainly comes from the direct coupling between the resonator and the substrate. In general, the
dissipation of the solid-state mechanical system is inversely proportional to the oscillator's mass. As the size decreases, it becomes a challenge to achieve very low
mechanical
dissipation. Efforts have been made to pursuit lower dissipation along different ways, such as employing novel improvement in material \cite{material1,material2}, geometry
\cite{geometry}, and strain \cite{Strain}. A dissipation down to the scale of $\rm ~mHz$ \cite{Strain} has been reported. For a ultra-light mechanical resonator made of
1D or 2D materials with a mass of $\rm  10^{-18}~$g, a dissipation of $\rm ~11 Hz$  is achieved \cite{nanotube}.

Different from solid-state mechanical systems, the dissipation of a levitated mechanical oscillator in ultra high vacuum is theoretically considered to be much lower
\cite{levitated theory}. In a recent experiment, a dissipation about $\rm 10 ~\mu Hz$ has been realized using optical levitation in high vacuum \cite{optical
levitation1,optical levitation2,optical levitation3}, and a dissipation as low as about $\rm 81 ~\mu Hz$ has been realized in an electric trap \cite{electric levitation}.
Typically, optical and electrical levitated oscillators work at room temperature due to the required power input, while on the other hand, magnetic levitated oscillators,
either Meissner levitated  \cite{Meissner1,Meissner2} and diamagnetic levitated \cite{diamagnetic1,diamagnetic2}, are fully passive with no energy inputs and so naturally
suitable for low temperature operation, and have recently been demonstrated to have mechanical dissipations at the order of magnitude of $\rm 10~\mu Hz$
\cite{diamagnetic1,diamagnetic2,Meissner2}. However, realization of a diamagnetic levitated oscillator operating at low temperature is still elusive
\cite{diamagnetic1,diamagnetic2,diamagnetic3,diamagnetic4}.

In this letter, we report a diamagnetic levitated micro-mechanical oscillator working at low temperature. The levitation is realized with a magneto-gravitational trap, where a
diamagnetic micro-sphere of mass of $\rm 540~pg$ (1~pg $\rm = 10^{-12}~$g) is stably levitated. We observed a ultra-low mechanical dissipation with damping rate $\gamma /2 \pi$
of $0.59~\pm 0.11~\mu $Hz at resonance frequency of $11.7\ $~Hz, the corresponding quality factor (Q) is $2\times10^7$. The reported mechanical dissipation is more than three
orders of magnitude lower than traditional solid-state oscillators \cite{material1,material2,geometry,Strain}, and one to three orders lower than reported optical levitated
oscillators \cite{optical levitation1,optical levitation2,optical levitation3}, Meissner levitated \cite{Meissner1,Meissner2} and electrical levitated oscillators
\cite{electric levitation}, and also about an order of magnitude improvement over previous experiments based on the same principle but working at room temperature
\cite{diamagnetic1,diamagnetic2,diamagnetic3,diamagnetic4}, the reported mechanical dissipation is even slightly smaller than the start-of-art milligram-scale pendulum
oscillator \cite{Pendulum1,Pendulum2}. The behavior of this system used as force and acceleration sensors is evaluated and its potential applications to realizing quantum
spin-mechanics systems are discussed.

\begin{figure}
\includegraphics[width=1.0\columnwidth]{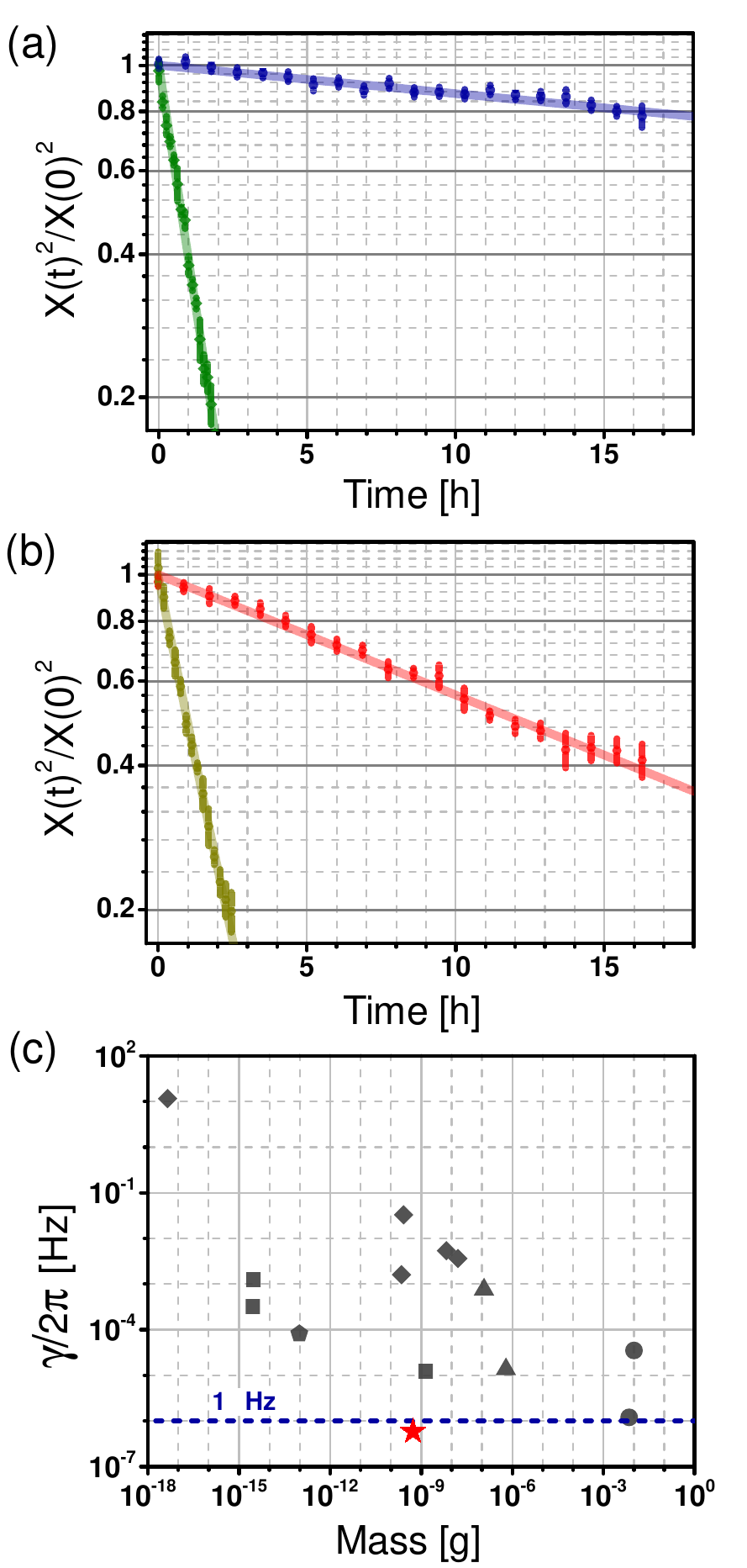}
\caption{(Color online). Measured mechanical dissipation.
(a), Logarithmic plot free ring-down of normalized oscillation energy  $ X(t)^2/X(0)^2$ coppresponding to oscillation mode with resonance frequency 11.7~Hz as function of time.
Green and blue dots are results with statistic errors measured under high purity helium gas and vacuum. Solid curve lines are fit to the exponential decay given by Eq.
\eqref{decay}
(b), Same as (a) but for the oscillation mode with resonance frequency 8.7~Hz, with dark yellow the helium gas and red the vacuum.
(c) The mechanical dissipation $\gamma/2\pi$ and the corresponding oscillator mass for a variety of mechanical systems based on different mechanisms. Red star is dissipation of
11.7~Hz mode in this experiment, rhombuses are solid-state oscillator \cite{material1,material2,Strain,geometry,nanotube}, squares are optical levitated oscillator
\cite{optical levitation1,optical levitation2,optical levitation3} , the pentagon is electrical levitated oscillator \cite{electric levitation}, triangles are Meissner
levitated oscillator \cite{Meissner1,Meissner2} and circles is pendulum oscillator with milligram-scale \cite{Pendulum1,Pendulum2}.
}
 \label{results}
\end{figure}

\textit{Experimental system.} \textbf{---}
The experimental setup is shown in Fig.~\ref{setup}. A magneto-gravitational trap (denoted as MGT hereafter) are placed in low-temperature and high-vaccum environment. The low
temperature is generated by a cryogen-free cryostat. To overcome the strong vibration generated by the operation of the cryostat, a special designed string-mass suspensions
system,  as shown in Fig.~\ref{setup} (a), is used to isolate vibration noise. The suspension structure consists of three stages, with each stage corresponds to a suspension
load mass as well as character frequency.  The designed isolation at frequency 8Hz is 54~dB and is expected to be better at resonance frequency of diamagnetic levitated
micro-mechanical oscillator (see Table~\ref{tab:table1} for parameter of suspension structure).
\begin{table}[h]
	\caption{\label{tab:table1} Designed parameters of the vibration isolation system. The characteristic frequency is the mass-spring suspension resonance frequency at
vertical direction ($z$ axis) which is significantly larger than the frequency in horizon ($x,y$), isolation give actuation in square of vibration amplitude response at
frequency 8Hz.  }
	\begin{ruledtabular}
		\begin{tabular}{cccc}
			Stage   & Mass (kg) & Character frequency (Hz)  & Isolation (dB)   \\ \hline
			 1st     &  7    & 1.4  & 29 \\ %\hline
			 2nd     &   1.6      & 2.5   & 19 \\
			 3rd     &   0.08      & 4.6   & 6 \\
		\end{tabular}
	\end{ruledtabular}
\end{table}

%Nano-positionors are used to move the position of magneto-gravitational trap so that the trap centre is maintain at the fouce of the detection optics during cooling process,

The magneto-gravitational trap is similar to the one used at room temperature \cite{diamagnetic2} but with modifications in geometry. It is generated by a set of SmCo magnets
with octagonal bilayer geometry as shown in Fig.~\ref{setup} (b). The oscillator is a microsphere of polyethylene glycol which is loaded into the trap by a home-built
nebulizer,
the charge on the micro-sphere is eliminated by following a standard procedure \cite{diamagnetic2}. After pumping the sample chamber into vacuum, the system is cooled down to
and
the sample chamber is then maintained at 3K during the experiment. The temperature of the magneto-gravitational trap, due to its  weak thermal connection to cold plate of
sample chamber, has a higher temperature about 7K.

To obtain the position and the dynamics of the levitated micro-sphere, a weak $633~$nm laser with power less than $50~\mu$w is sent into the chamber via single mode fiber and
is
loosely focused on the microsphere, and the scattered light from the microsphere is collected vertically via a lens and comes out of the cryostat via windows. To realize
feedback cooling and excitation of the micro-sphere, an avalanche photodiode detector (APD) is firstly used to measure the intensity of scattered light that contains the real
time information of the microsphere position, then the electrical signal of the detector is amplified and sent to a computer where its amplitude and phase corresponding to the
oscillator resonance frequency are calculated, and a program based feedback circuit is used to control the  arbitrary waveform generator which generated the feedback signal.
Typically, the feedback bandwidth is of the scale of 0.1~Hz which is fast enough for our system. The output electric signal goes back to the sample stage via a small coil near
the micro-sphere and generating a weak magnetic field. Combined with the magnetic field gradient of the MGT finally, it becomes a feedback force applied on the microsphere. Due
to the limitation of the dynamic range of APD, we used a CMOS camera instead so that the absolute position of the microsphere was directly recorded. Fig.~\ref{setup} (c) and
(d) is typical images of micro-sphere excited motion corresponding to two resonance modes.

\textit{Measurement method and results.} \textbf{---}
To obtain the mechanical dissipation of the micro-oscillator, we start from the equation of motion of the a micro-oscillator of math $m$ of one oscillation mode with a
resonance frequency of $\omega_0 /2\pi)$:
\begin{equation}
\label{EOM} m\ddot{x}+m \gamma \dot{x}  + m \omega^2_{0} x + m \epsilon x^3= F_{\rm fluc}(t),
\end{equation}
where the second term describes the dissipation the oscillator with a damping rate $\gamma / 2 \pi$, the fourth term is the Duffing nonlinearity, and  $F_{\rm fluc}(t)$ is the
total fluctuation force, including thermal fluctuation, external vibration etc. The motion can be written in the form $ x(t) = X(t)~\rm cos(\omega_0 t+ \varphi (t)) $ with
$X(t)$ and $\varphi(t)$ the slow-varying amplitude and phase, respectively. There are several commonly used experimental methods to determine the dissipation $\gamma$. One is
to measure the system's frequency response. For
harmonic oscillator, the corresponding linewidth equals to $\gamma/\pi$, but in the presence of nonlinearity, the measured linewidth is broadened significantly, even when it
is driven by solely the thermal fluctuation \cite{optical levitation1,diamagnetic2}. We obtained $\gamma$  by energy autocorrelation proportional to $\langle X(t)^2 X(0)^2
\rangle$, with $X(t)$, as it has been shown that energy autocorrelation is insensitive to nonlinearity \cite{Rugar1}. In the actual realization, we directly measure the free
decay of $ X(t)^2 $, to do this we first excited the oscillator to initial amplitude $X(0)$ that is much higher (more than 10 times) than the background stochastic motion
amplitude $\langle X_{\rm ba} \rangle ^2$ driven by the fluctuation force $F_{\rm fluc}(t)$, and the theoretical evolution can be written as:
\begin{equation}
\label{decay} X(t)^2 =  X(0)^2 e^{-t\gamma}
\end{equation}
with error less than a few percent  \cite{Rugar1,diamagnetic2}.
Under our experimental conditions, $X(t)^2$ were obtained using a CMOS camera based on the method in Ref.[37]. The position of the microsphere was recorded for a duration so as
to obtain  an average of $X(t)^2$ and its statistical deviation. The measured data were then fitted with Eq. (4) to give the dissipation rate $\gamma$. Then the dissipation
rate is obtained by fitting the measured data with Eq. (\ref{decay}).

The diameter of the micro-sphere used in this experiment is $9.8 \pm 0.9~\mu$m, which is obtained in advance from the dissipation $\gamma$ measured at high purity helium gas
environment with pressure $2 \times 10^{-6}~$mbar. As the helium gas at a temperature of 3K can still be considered as a idea gas, the diameter is simply related to the
microsphere oscillation damping rate and the pressure as a result of background gas collisions \cite{levitated theory}: $ \gamma = (16/\pi)(P/ \nu R \rho) $, with
$P$ and $\nu$ the pressure and the mean speed of the background gas, respectively, and $\rho$ the density of the micro-sphere. Then we turned off
the helium gas and pumped the chamber to high vacuum with background pressure $3.3 \times 10^{ -7} $ mbar (measured at room temperature), which is maintained stable during the
experiment. Fig.~\ref{results} (a) and (b) shown the normalized damping curve as $X(t)^2 /X(0)^2$, and the results is summarized in Table~\ref{tab:table2}, for the vibrational
mode of resonance frequency 11.7~Hz (mode 1) , we observed a ultra-long damping time $\tau = 1/\gamma$ of $2.7\times 10^{5}~$s (about 70 hours), corresponding to a  mechanical
dissipation of $\gamma/2\pi = 0.59 \pm 0.11\rm ~\mu$Hz and a quality factor $Q = 2.0\pm 0.4 \times  10^7$. For the vibrational mode of resonance frequency 8.7~Hz (mode 2), the
mechanical dissipation is $\gamma/2\pi = 2.6 \pm 0.1 \rm ~\mu$Hz, about 4 times larger than that of mode 1. A possible origin of such a difference may come from
boundary effects of residual helium gas \cite{boundary} in carrying out our experiment. Fig.~\ref{results} (c) plots the comparison of our results with currently
reported micro- and nano-mechanical using different principle. It clearly shows that our system performs with significant lower dissipation in a wide range of mass, and is even
comparable to those milligram-scale mechanical system whose mass is larger by about 7 orders of magnitude \cite{Pendulum1,Pendulum2}.

\begin{table}[h]
	\caption{\label{tab:table2} Measured mechanical dissipation at low temperature and high vacuum. The diameter of the oscillator is $9.8 \pm 0.95~ \mu$m, and the
corresponding mass is $\rm 540 \pm 160~ p$g. The errors show the statistic deviation at the $95~\%$ confidence level.}
	\begin{ruledtabular}
		\begin{tabular}{cccc}
			 Mode    & $\omega_0 / 2 \pi$ (Hz) & $\gamma/ 2 \pi$ ($\rm \mu Hz$)  & Q ($10^7$) \\ \hline%\hline
			 1     &  11.7      & $0.59\pm 0.11  $   & $2.0\pm 0.4 $ \\ %\hline
             2     &  8.4       & $2.6\pm 0.1  $  & $0.34\pm 0.02 $ \\
		\end{tabular}
	\end{ruledtabular}
\end{table}

\begin{figure}
\includegraphics[width=1.0\columnwidth]{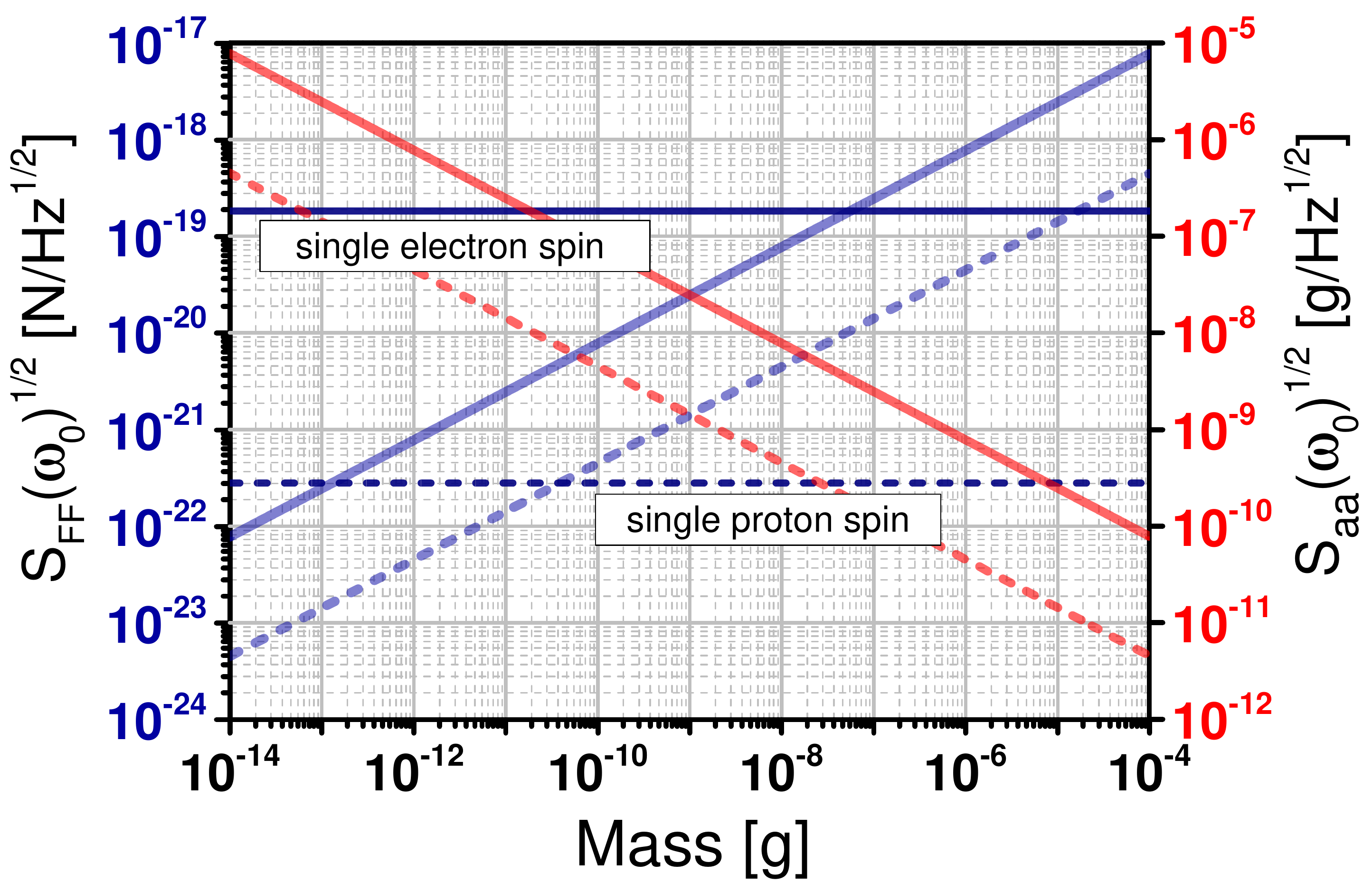}
\caption{(Color online). Estimation of performance of force and acceleration sensing.  Left axis, force sensitivity as function of mass at environment temperature of 3K (light
blue solid line ) and 10~mK (light blue dash line), force from single electron spin (dark blue solid line) and single proton spin (dark blue dash line) are shown for
comparison.
Right axis, same as left but of acceleration sensitivity. The experimently measured dissipation $\rm \gamma/2 \pi = 0.59 ~\mu Hz$ and the magnetic gradient $\rm G =
10^4~T/m$ are adopted in the estimation.
}
 \label{Discussion}
\end{figure}

\textit{Discussion and summary.} \textbf{---}
The potential performance of our diamagnetic-levitated oscillator used for force and acceleration sensing can be evaluated from the dissipation achieved in our experiment. The
sensitivity of force sensor is given by Eq. \eqref{force}, and the sensitivity of acceleration is characterized by noise power density $S_{\rm aa}(\omega_0) = 4\gamma k_B T/m$,
which differs from that of the force in that large mass is advantageous. In principle, a magneto-gravitational trap can levitate particles of a very wide range of mass of the
order of magnitude of $\rm ~ 10^{-14}$
\cite{diamagnetic2} to $~10^{-3}~$g  \cite{diamagnetic3, diamagnetic4}.  Fig.~\ref{Discussion} shows the force and acceleration at 3K which corresponding to this experiment and
10~mK which can be reached by a commercial dilution refrigerator. At 3K, we see that the force sensitivity is enough to detect a single electron when the oscillator's mass is
smaller than $4\times10^{-9}~$g. In the estimation, a magnetic gradient $G = 10^4$~T/m from a typical  magneto-gravitational trap  \cite{diamagnetic1,diamagnetic2} is adopted.
A single proton spin is also detectable when the the oscillator's mass is smaller than $100$~fg. On the other hand, for a mass of $\rm 10^{-6}$~g, the acceleration sensitivity
reaches $\sqrt{S_{\rm aa}(\omega_0)} < 10^{-10}g/ \rm \sqrt{Hz}$ ($g = 9.8~{\rm m/s^2}$) at 3K, comparable to the state-of-art method reported so far \cite{acceleration}.

The potential of application of this system to the study of macroscopic quantum process can be estimated by assuming an oscillator resonance frequency $\omega_0 / 2\pi \simeq
100$~Hz \cite{diamagnetic1} and at a $10$ mK environment temperature. We find that $\omega_0 Q/ 2\pi \simeq 10^{10}$~Hz $\gg k_BT/(2\pi\hbar) $ and Eq.~\eqref{quantum} is well
satisfied. Therefore, the quantum coherence of the system could last for many oscillation periods.
Especially, it is interesting to consider applying the system to the realization of quantum spin-mechanics dynamics \cite{spin-mechanics1,spin-mechanics2}.  Considering the
interaction between electron spin, the spin-mechanical coupling strength $\lambda = Gx_{\rm zpl}\mu_{e}/\hbar$, where $x_{\rm zpl} = \sqrt{\hbar/m \omega_0}$ is the zero point
motion and $\mu_{e}$ is the electron spin magnetic moment, for oscillator with mass of $10^{-13}$~g, we have $\rm \lambda/2\pi \simeq 10~kHz$, which is four orders of magnitude
larger than the decoherence rate  $1/T_2$ of electron spin at low temperature \cite{spin-T2}, and four orders of magnitude larger than thermal decoherence rate
$\gamma_{\rm th}$ of the oscillator.  The coopertivity here is $C=\lambda^2 T_2 Q  \hbar  /( k_{B}T) \simeq 10^8$, which marks the onset of highly coherent quantum effects.

In summary, we report the realization of a cryogenic diamagnetic-levitated micro-mechanical system with a recorded low dissipation value, which provides a new way to achieve
high-performance sensing and quantum control of macroscopic mechanical dynamics.

%\begin{figure}[hbtp]
%\centering
%\includegraphics[width=1\columnwidth]{figure2.pdf}
%\caption{(Color online). The drive-response properties of resonator T by monitoring resonator D.
%(a), The response of T as a function of the drive frequency $\omega_{\rm dr}$ measured by the displacement of D.
%The pump frequency $\omega_{\rm pu}$ was tuned to $\omega_{\rm pu} %=\omega_{\rm d} - \omega_{\rm dr}$ when $\omega_{\rm dr}$ was swept.
%(b), The optically measured thermal noise spectrum under the same drive condition as (a).
%(c), Same as (a), but under different a.c. pump voltage. The fact that no peak being observed
%when the a.c.\ pump is tuned off resonance ascertains that the measured peak is from resonator T.
%(d), The displacements $\Delta x_d$ %for resonator D versus $\Delta x_t$  for
%resonator T %under a fixed pump voltage, with red line being a linear fitting.
%The black curves in the sub-figures (a-c) are fits of Lorentz shape to guide eyes.
%}
% \label{fig2}
%\end{figure}

We thank Zhujing Xu, Fei Xue and Jie Zhao for helpful discussion. This work was supported by the National Key R\&D Program of China (Grant No. 2018YFA0306600), the National
Natural
Science Foundation of China (Grant No.~61635012, No.~11675163, No.~11890702, No.~81788101, No.~11761131011, and No.~11722544),  the CAS (Grant No. QYZDY-SSW-SLH004, Grant No.
GJJSTD20170001), the Fundamental Research Funds
for the Central Universities (Grant No. 021314380149), and the Anhui Initiative in Quantum Information Technologies (Grant No. AHY050000).
Z.-Q. Yin is supported by National Natural Science Foundation of China under Grant No. 61771278 and Beijing Institute of Technology Research
Fund Program for Young Scholars.

\end{document}